# Simulation and Prediction of Countercurrent Spontaneous Imbibition at Early and Late Time Using Physics-Informed Neural Networks

Jassem Abbasi, Pål Østebø Andersen*

**Department of Energy Resources, University of Stavanger, 4021 Stavanger, Norway**

* pal.andersen@uis.no

## Abstract

The application of Physics-Informed Neural Networks (PINNs) is investigated for the first time in solving the one-dimensional Countercurrent spontaneous imbibition (COUCSI) problem at both early and late time (i.e., before and after the imbibition front meets the no-flow boundary). We introduce utilization of Change-of-Variables as a technique for improving performance of PINNs. We formulated the COUCSI problem in three equivalent forms by changing the independent variables. The first describes saturation as function of normalized position X and time T; the second as function of X and $Y=T^{0.5}$; and the third as a sole function of $Z=X/T^{0.5}$ (valid only at early time). The PINN model was generated using a feed-forward neural network and trained based on minimizing a weighted loss function, including the physics-informed loss term and terms corresponding to the initial and boundary conditions. All three formulations could closely approximate the correct solutions, with water saturation mean absolute errors around 0.019 and 0.009 for XT and XY formulations and 0.012 for the Z formulation at early time. The Z formulation perfectly captured the self-similarity of the system at early time. This was less captured by XT and XY formulations. The total variation of saturation was preserved in the Z formulation, and it was better preserved with XY- than XT formulation. Redefining the problem based on the physics-inspired variables reduced the non-linearity of the problem and allowed higher solution accuracies, a higher degree of loss-landscape convexity, a lower number of required collocation points, smaller network sizes, and more computationally efficient solutions.

**Keywords:** Spontaneous Imbibition; Physics Informed Neural Networks; Partial Differential Equations, Change of Variables

## 1- Introduction

Spontaneous imbibition (SI) is a flow phenomenon where a wetting phase diffuses into porous media and displaces a non-wetting phase due to capillary forces. This process commonly occurs in geological problems, such as soil water infiltration, hydrocarbon production, and carbon sequestration. Other important fluid displacement mechanisms in porous media include gravity and advection. Analyzing the spontaneous imbibition process provides insights into the wetting condition of the system (if the surrounding fluid wets the porous system) and the governing capillary pressure curve regarding how strong capillary forces are present and how much fluid displacement they can cause [1]. The boundary conditions influence the flow behavior during SI. In systems where both fluids access the system through the same surface, they flow counter-currently. This is called counter-current SI (COUCSI). When some system surfaces are exposed to the wetting phase and other surfaces to non-wetting phase, the fluids predominantly flow in the same direction, which is called co-current SI (COCSI) [2].

During COUCSI, we distinguish between the early time (ET) period (before the invading saturation profile has reached the closed boundary) and the late time (LT) period (after the no-flow boundary has been reached). In ET, the advancement of the front is proportional to the square root of time [3], and there is a self-similarity behavior in the saturation front of fluid versus a similarity variable [4,5]. The recovery stays proportional to the square root of time during ET. However, it has been demonstrated comprehensively that recovery can follow this trend long into the LT regime [6–8]. The explanation is that the imbibition rate at the inlet, which controls recovery, acts as if it is still ET until the saturation profile has changed sufficiently from the self-similar ET saturation profile at the inlet [9].

The COUCSI problem can be solved numerically with finite differences (FD) approaches using, e.g., reservoir simulators [10,11]. Analytical solutions can be derived for very specific saturation functions or piston-like displacement [12–14]. Semi-analytical solutions were derived by McWhorter & Sunada [15] that hold for any input parameters and saturation functions, except being limited to ET. Bjørnarå & Mathias [16] noted numerical challenges in evaluating their solution and proposed more stable schemes. Numerical approaches are usually required for more complex problems involving more general geometries or changing boundary conditions. Schmid & Geiger [17] used McWhorter and Sunada's solution to propose a universal time scale for COUCSI. Andersen [8] showed that all one-dimensional (1D) COUCSI problems (regardless of wettability, saturation functions, and other parameters) could be normalized to the same form, which only depends on a normalized diffusion coefficient with a mean equal to one. By correlating the coefficient shape with recovery behavior, ET and LT recovery were predicted accurately without the need for numerical methods.

Physics-informed neural networks (PINNs) provide an alternative approach to solving ordinary- and partial- differential equations (ODEs and PDEs) that govern various engineering problems [18]. PINNs have recently been applied to mathematical analysis of flow in porous media: Fraces & Tchelepi [19], Almajid & Abu-Al-Saud [20], and Rodriguez-Torrado et al. [21] utilized PINNs for solving the Buckley-Leverett (BL) equation describing the displacement of oil by water flooding. A key challenge was capturing the shock front behavior. Deng & Pan [22] applied PINN to study 1D spontaneous imbibition using Lagrangian formulation. They investigated ET flow for self-similar counter- and co-current flow and also considered more general boundary conditions during co-current SI where the ET flow would not be self-similar due to countercurrent production at the inlet. In the self-similar system, they could determine the saturation-dependent characteristic flow function, fluxes as a function of time, and solutions of saturation along the core. The sensitivity of problem formulation of COUCSI to capture self-similarity and other properties has not been explored. To our knowledge, the current work is only the second study investigating PINN applied to (COUC)SI. A problem's non-linearity may reduce the accuracy of the PINNs solutions [23,24]. A series of methodologies, such as singular value decomposition [25], convolutional autoencoders [26], and Fourier transformation [27] have been suggested for tackling the dimensionality issues by transforming the network operations to lower frequency spaces. However, this issue is still an open question that needs to be addressed [28]. We propose that a proper mathematical formulation of the system will have impact on the non-linearity and, thus, the ability of a PINN to represent the solution accurately. It is noted that defining features based on physics is a meaningful and established way of constraining machine learning models [29,30]. However, the role of modifying the independent variables while representing the same PDE problem has not been explored and not in a PINN context.

This work investigates whether a PINN-based approach can effectively solve the 1D COUCSI problem - for the first time at both ET and LT periods - while only using the same input as an FD numerical simulator; hence, no utilization of labeled training data from other sources (such as experiments, analytical or numerical solutions) to guide the PINNs training. We compared PINNs to FD approach in terms of accuracy and computational run-time. Additionally, we introduced a novel improvement by examining the impact of reformulating the COUCSI PDEs based on different independent variables, derived from the system's physical characteristics, on PINN accuracy, network size, and collocation points. We also evaluated to which level the PINN solutions from different formulations preserve physical properties of the COUCSI system, such as self-similarity, Total Variation (TV) preservation, and square root of time recovery behavior.

In the following section, we present the mathematical theory behind COUCSI and its formulations, the PINN methodology, and its implementation in this work. The results of applying PINN to solve the COUCSI problem for the different formulations for different test cases and different systematic training strategies are then presented. Afterwards, we apply PINNs to regenerate a set of experimental tests. The work is finalized with conclusions.

## 2- Mathematics of Spontaneous Imbibition

We investigate a two-phase flow system with immiscible and incompressible fluids. The porous medium is incompressible and homogeneous. The flow occurs in the horizontal direction without gravitational forces. All the faces are closed except the face at $x = 0$, where it is exposed to the wetting phase. The saturation of the wetting phase at the open face is fixed to $s_w^{eq}$, defined by a zero capillary pressure $P_c(s_w^{eq}) = 0$. The system is initially saturated with the non-wetting phase. The stated conditions result in 1D COUCSI, as shown in Figure 1. Darcy's law in 1D gives Darcy velocity ($u_i$) by writing:

$$u_i = -\lambda_i [\partial_x p_i], \qquad \lambda_i = \frac{K k_{ri}}{\mu_i}, \qquad (i = w, nw), \qquad 1$$

where $\lambda_i$ is mobility, $p_i$ is pressure, $K$ and $k_{ri}$ are absolute permeability and relative permeability, respectively, and $\mu_i$ is viscosity. The index $i$ refers to phase-specific properties for the wetting and non-wetting phases ($w$, $nw$). The conservation law determines the volumetric transport of phases in time:

$$\phi \partial_t(s_i) = -\partial_x(u_i), \qquad (i = w, nw), \qquad 2$$

where $s_i$ is phase saturation, and $\phi$ is porosity. The saturations are constrained by volume conservation ($s_w + s_{nw} = 1$), and the positive capillary pressure function constrains the pressures ($p_{nw} - p_w = P_c(s_w)$). Combining these assumptions, we obtain the capillary diffusion equation describing 1D countercurrent spontaneous imbibition in the form of a second-order, elliptic PDE:

$$\phi \partial_t(s_w) = -K \partial_x(\lambda_{nw} f_w \partial_x P_c), \quad (0 < x < L) \qquad 3$$

where $f_w = \lambda_w / (\lambda_w + \lambda_{nw})$. The implementation in Eq. 3 reduces the problem's dimensionality compared to Eq. 1 by removing the phase pressures and constraining the equations to a single PDE. The boundary and initial conditions are given by:

$$s_w(x = 0, t) = s_w^{eq}, \qquad \partial_x s_w|_{x=L} = 0, \qquad s_w(x, t = 0) = s_{wr} \qquad 4$$

representing a fixed saturation at the inlet corresponding to zero capillary pressure, a closed boundary at $x = L$, and a uniform initial saturation. $s_{wr}$ is also the residual water saturation. Considering $s_{nwr}$ as the irreducible non-wetting phase saturation, when $s_w^{eq} = 1 - s_{nwr}$ (oil is immobile) the system is strongly water-wetting, while if $s_{wr} < s_w^{eq} < 1 - s_{nwr}$, the system is considered mixed-wet.

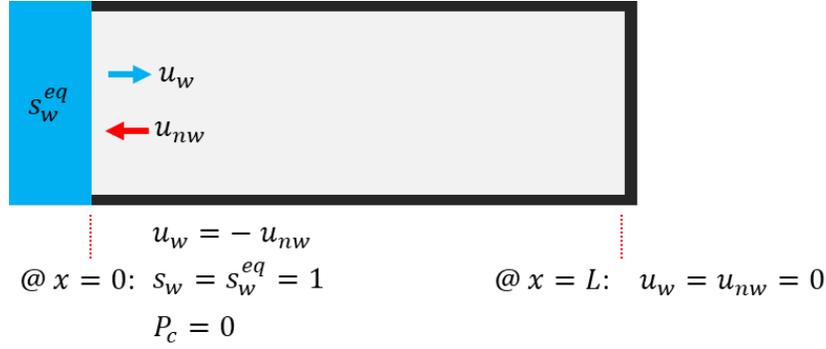

*Figure 1: The COUCSI geometry and the applied boundary conditions. The core is initially saturated by the non-wetting phase and exposed to the wetting phase at $x = 0$. Other faces are closed to flow.*

### 2-1- Scaled Representations

Three different scaled representations will be considered of the flow equations (Eqs. 2 or 3). The Change-of-Variables technique was used, which is defined as the substitution of the variables of a PDE by new variables that are a function of the original variables [31]. Each representation has different properties and hence potential advantages related to producing accurate solutions with PINN, which will be investigated.

### 2-1-1- Normalized Diffusion Equation (XT formulation)

By defining normalized spatial and temporal variables $X$, and $T$, and saturation $S_n$ the system of Eq. 3 and 4 can be written (see section A1 of the supplementary materials for details) into an equivalent dimensionless form [8]:

$$\partial_T S_n = \partial_X(\Lambda_n(S_n)\partial_X S_n), \quad (0 < X < 1) \qquad 5$$

$$S_n(X = 0, T) = 1, \quad \partial_X S_n|_{X=1} = 0, \quad S_n(X, T = 0) = 0 \qquad 6$$

The interval $0 < S_n < 1$ is where imbibition occurs. $\Lambda_n$ is a scaled diffusion coefficient with a mean one over the imbibing saturations. This scaled representation captures all wetting states and input parameter combinations and only depends on the shape of $\Lambda_n(S_n)$. Eq 5 and 6 are called the XT-formulation. This formulation is valid in full-time (FT) period, i.e., both ET and LT.

### 2-1-2- Square Root of Time Full-time Solution (XY-formulation)

By replacing the temporal variable from T to $Y = T^{0.5}$ and use it together with $X$, i.e., $S_n = S_n(X, Y)$, we can develop a system that is valid for both ET and LT flow periods. When reformulating Eq. 5, we use $Y$ in the spatial derivatives [9]:

$$\partial_Y S_n = 2Y\partial_X(\Lambda_n(S_n)\partial_X S_n), \qquad 7$$

$$S_n(X, Y = 0) = 0, \quad S_n(X = 0, Y) = 1, \quad \partial_X S_n|_{X=1} = 0 \qquad 8$$

We refer to this system as the XY formulation.

### 2-1-3- Self-similar Early-time Solution (Z-formulation)

If we consider the time before the no-flow boundary is encountered, the 1D COUCSI system behaves in a self-similar way. The role of position and time can be combined into one variable $Z = X/T^{0.5}$ which determines the solution for a given function $\Lambda_n$, i.e., $S_n = S_n(Z)$. Using this variable, the system of Eq. 5 and 6 can be expressed as [9]:

$$Z \partial_Z S_n = -2 \partial_Z (\Lambda_n(S_n) \partial_Z S_n), \qquad (0 < Z < \infty) \qquad 9$$

$$S_n(Z = \infty) = 0, \qquad S_n(Z = 0) = 1 \qquad 10$$

with the exception that the closed boundary condition in Eq. 6 is replaced by obtaining the initial condition at an infinite distance. We call Eqs. 9 and 10 the *Z*-formulation.

### 2-2- Total Variation (TV)

The Total Variation (TV) concept in mathematics is the magnitude of the arclength of a curve related to a continuous function in an interval of interest [32]. It can represent the amount of oscillation in a numerical solution and whether it increases with time; hence it is a valuable concept in numerical stability analysis [33]. In this context, we consider the Total Variation in saturation along the core as a function of time:

$$TV(S_n(\cdot, t)) = \int_{X=0}^{1} |\partial_X S_n| \, dX = \int_{X=0}^{1} |dS_n| \qquad 11$$

In other words, TV measures the total change in saturation along the core. As we have scaled the system to have a saturation equal to one at the inlet and to be zero initially, the TV is expected to be one at the ET. However, at LT, the saturation increases from zero at the closed boundary and hence reduces the TV. We thus expect our PINN solution to produce $TV = 1$ at the early time and $TV$ declining from one towards zero (when the saturations are uniformly equal to 1) at the late time.

### 2-3- Physics-Informed Neural Networks

Physics-informed Neural Networks (PINNs) are a subset of scientific deep learning that try to solve PDEs/ODEs using a data-driven approach [34]. If $S_n(X,T)$ represents the solution of a 1D COUCSI system based on $XT$ formulation, the PINNs model is defined as:

$$S_n(X,T) + \mathcal{N}[S_n; \theta] = 0; \quad x \in \mathbb{R}, t \in \mathbb{R} \qquad 12$$

Where $\mathcal{N}[S_n; \theta]$ is the equivalent differential operator describing the investigating problem, and $\theta$ is its constitutive (trainable) parameters, including the network's weights and biases. The PINNs model is developed by constraining $\mathcal{N}[S_n; \theta]$ during the training, using the PDE representing the physical system and the constraining initial or boundary conditions. The training is carried out by utilizing automatic differentiation (AD) while back-propagation. By applying the successive chain rule, automatic differentiation computes gradients with an efficient computational cost.

The physical specification of the system is imposed on the PINN by the definition of a series of loss functions representing the mean squared errors (MSE) in the residuals of the system's underlying PDE ($\mathcal{F}$), initial conditions ($\mathcal{I}$), and boundary conditions ($\mathcal{B}$):

$$\mathcal{L}_\mathcal{F}(\theta) = \int_\Omega |\mathcal{F}(\mathcal{N})|^2 \, dx \qquad 13$$

$$\mathcal{L}_\mathcal{I}(\theta) = \int_\Omega |\mathcal{I}(\mathcal{N})|^2 \, dx \qquad 14$$

$$\mathcal{L}_\mathcal{B}(\theta) = \int_\Omega |\mathcal{B}(\mathcal{N})|^2 \, dx \qquad 15$$

where $\mathcal{L}_\mathcal{F}$ denotes the loss due to deviation from the governing PDE in the domain $\Omega$. Also, $\mathcal{L}_\mathcal{I}$ and $\mathcal{L}_\mathcal{B}$ are the losses related to the initial and boundary conditions, respectively. $\mathcal{L}_\mathcal{B}$ includes the conditions at both faces, i.e., $X = 0$, and $X = 1$ (except for Z formulation that is only valid at ET and only has a boundary condition at $X = 0$). If observation points (from sources like experiments or analytical/numerical solutions) are also available, an additional loss term representing the deviations from the true values may be defined. However, in this work, since no observation data is used in the training of the model, we neglected this loss term. Consequently, the total loss function is the weighted summation of all the defined losses:

$$\mathcal{L}_t(\theta, x) = \omega_1 \mathcal{L}_\mathcal{F}(\theta) + \omega_2 \mathcal{L}_\mathcal{I}(\theta) + \omega_3 \mathcal{L}_\mathcal{B}(\theta) \qquad 16$$

Where $\alpha_i$ is the weight factor for each loss term. Except for $\omega_3$, for which we set the value to 0.10, all other loss weight factors were assigned unity values. The above definition can also be defined for other forms of the COUCSI PDEs, i.e., *XY* and *Z* formulations. Based on the parameters of the applied formulation, we can define the PDE-, initial condition-, and boundary condition- residuals as below:

*XT* Formulation:

$$\mathcal{F}(X, T) = \partial_T S_n - \partial_X (\Lambda_n(S_n) \partial_X S_n) \qquad 17$$

$$\mathcal{I} = S_n(X, T = 0), \quad \mathcal{B}_1 = S_n(X = 0, T) - 1, \quad \mathcal{B}_2 = \partial_X S_n \qquad 18$$

*XY* Formulation:

$$\mathcal{F}(X, Y) = \partial_Y S_n - 2Y \partial_X (\Lambda_n(S_n) \partial_X S_n) \qquad 19$$

$$\mathcal{I} = S_n(X, Y = 0), \quad \mathcal{B}_1 = S_n(X = 0, Y) - 1, \quad \mathcal{B}_2 = \partial_X S_n \qquad 20$$

*Z* Formulation:

$$\mathcal{F}(Z) = \partial_Z S_n + \frac{2}{Z} \partial_Z (\Lambda_n(S_n) \partial_Z S_n) \qquad 21$$

$$\mathcal{I} = S_n(Z = \infty), \quad \mathcal{B}_1 = S_n(Z = 0) - 1, \quad \mathcal{B}_2 = \partial_Z S_n \big|_{(Y, Z = \frac{1}{Y})} \qquad 22$$

$\mathcal{B}_2$ in *Z* formulation is theoretically trivial (in the ET period) and not obligatory in the training process; however, we calculated it at the deployment stage for quality check.

During our observations, we noticed that $\mathcal{F}$ values near the flowing boundary were initially high, which had a significant impact on the PINN solutions. To address this, we omitted the $\mathcal{F}$ values (in Eqs. 17 and 19) in the $X < 0.01$ range, during the training process.

## 3- Problem Setup
### 3-1- 1D COUCSI cases

We considered two 1D COUCSI cases (A and B) with different frontal behavior. The rock/fluid properties were identical in both cases, except for their saturation functions. The capillary pressure curve for case A is generated based on Kumar et al. [35]. The saturation curves in case B are artificially generated to create a steeper saturation profile, different than case A. The main goal of having two cases is to investigate the capabilities of PINNs in capturing various flow behaviors. The basic rock and fluid properties are shown in Table 1. Also, the utilized relative permeability and capillary pressure curves are shown in Figure 2. Figure 2c shows the scaled capillary diffusion coefficient curves where the differences in the saturation functions could generate two different trends. We expect differences in the flow regimes due to the differences in the scaled capillary diffusion curves. In this work, we utilized Corey-type relative permeabilities [36],

$$k_{rw} = k_{rw}^*(S)^{n_w}, \qquad k_{rnw} = k_{rnw}^*(1-S)^{n_{nw}}, \qquad 23$$

Here, $n_w$ and $n_{nw}$ are Corey exponents and $k_{rw}^*$ and $k_{rnw}^*$ are the relative permeability endpoints. Also, S is the normalized water saturation as:

$$S = \frac{S_w - S_{wr}}{1 - S_{or} - S_{wr}} \qquad 24$$

The correlation introduced by Andersen et al.[37] is used for the capillary pressure curves:

$$P_c = \frac{a_1}{(1+k_1 S)^{n_1}} - \frac{a_2}{(1+k_2(1-S))^{n_2}} + a_3 \qquad 25$$

The parameters $a_1, a_2, a_3, k_1, k_2, n_1$, and $n_2$ determine the correlation shape. In case A, $a_1, a_2$, and $a_3$ are 15.52, 0.74, and 0.50 kPa, respectively, while in case B, $a_1, a_2$, and $a_3$ are 11.17, 5.17, and 5.00 kPa, respectively. Other parameters corresponding to the capillary pressure curves are similar and are presented in Table 1. The reference solutions for comparison with the PINN solutions were generated using a previously developed and verified core scale simulator IORCoreSim [38] for numerical simulation of two-phase flow in porous media. The core was split equally into 400 grid cells, while the final scaled simulated time (35 hrs for case A and 10 hrs for case B) was divided into 120 report steps: the first 60 were equally spaced in square root time by $dt^{0.5} = 0.0167$ hr$^{0.5}$ and the next 60 by 0.0819 hr$^{0.5}$ for case A, and similarly $dt^{0.5} = 0.0167$ hr$^{0.5}$ and 0.0360 hr$^{0.5}$ for case B.

**Table 1: Rock and fluid properties of the studied cases.**

| Parameters | Values | Parameters | Values | Parameters | Value |
|---|---|---|---|---|---|
| $K$ | 243.2 md | $k_{rw}^*, k_{rnw}^*$ | 1.0, 1.0 | $k_1$ | 7 |
| $\phi$ | 0.2 | $n_w, n_{nw}$ | 1.8, 1.4 | $k_2$ | 0.4 |
| L | 0.06 cm | $S_{wr}$ | 0.0 | $n_1$ | 2 |
| $\mu_w, \mu_{nw}$ | 1.0, 1.0 cP | $S_{or}$ | 0.0 | $n_2$ | 10 |

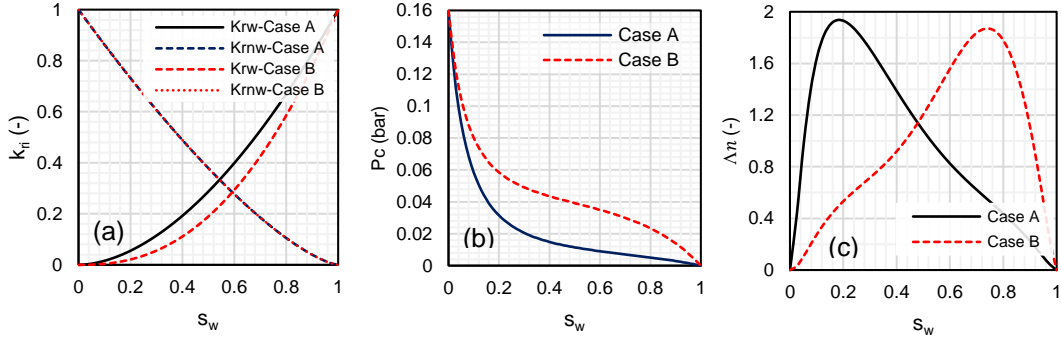

*Figure 2: The applied saturation functions for two generated cases (A and B). (a) relative permeability, (b) capillary pressure, (c) scaled capillary diffusion coefficient $\Lambda_n$. The capillary pressure curve of case A is from Kumar et al.* [35]. *Case B was artificially generated for comparison.*

### 3-2-  Network and Training Setup

For each of the introduced formulations (i.e., $XT$, $XY$, $Z$), we used a fully connected feed-forward multilayer perceptron (MLP) network, called the latent network, sandwiched between an encoder and a decoder. The network should output $S_n$ based on the corresponding inputs of each formulation, i.e., $S_n(X,T)$, $S_n(X,Y)$ or $S_n(Z)$. A schematic of the network setup is provided in Figure 3. The encoder is composed of an MLP with two linear layers followed by a normalization layer [39] and the decoder consists of an MLP with three linear layers. In both the encoder and decoder, the layer widths were the same as the width of the latent network, which was 50 nodes. The role of the encoder is to transform the dimension of the PINN's input to the width of the latent network by matrix multiplication, and the decoder maps the latent network output to the expected PINN output dimension. The latent network is composed of an MLP with five hidden layers with the *tanh* activation function and a width of 50 nodes. Fourier and inverse Fourier transform operations are performed on the input and output of the latent network, as discussed by Li et al. [40]. Performing the calculations in Fourier space improved the PINN performance and was applied as it is known to resolve deficiencies of standard PINNs in capturing high-frequency elements of the problem [41]. In the output of the PINN, we set a rectifier function to make sure $0 \leq S_n \leq 1$.

Based on the mathematical definition of the investigated formulation, the model may have one input (for the $Z$ formulation) or two inputs (for $XT$, and $XY$ formulations). However, all the formulations used the same network architecture/size to simplify comparisons. A relatively large model was chosen to provide enough network complexity to capture the solution for all formulations. A sensitivity analysis of network size is carried out later to examine the possibility of simpler networks.

The number of collocation points in the spatial dimension ($n_X$ or $n_Z$) were chosen based on one hundred equally spaced points. In the temporal dimension, the number of collocation points ($n_T$) was 45 for both cases, in the FT simulations; approximately 1/3 of the points were in the ET region, and the remaining were in the LT region. The temporal points were increasing by constant DY values, up to the total time. The model was trained by consolidating all collocation points into a singular batch (fully batched), resulting in a single backpropagation calculation per epoch.

The trainable parameters in the network were initialized using a uniform random scheme. The gradient-based Adam optimizer [42] was used for training the network with the starting learning rate 1e-4 and weight-decay 1e-5. We gradually reduced the learning rate by one order of magnitude during the training. The training was continued until the reduction in the loss term stopped. In all cases, the training could be terminated after $n_{ep}$=4000 epochs.

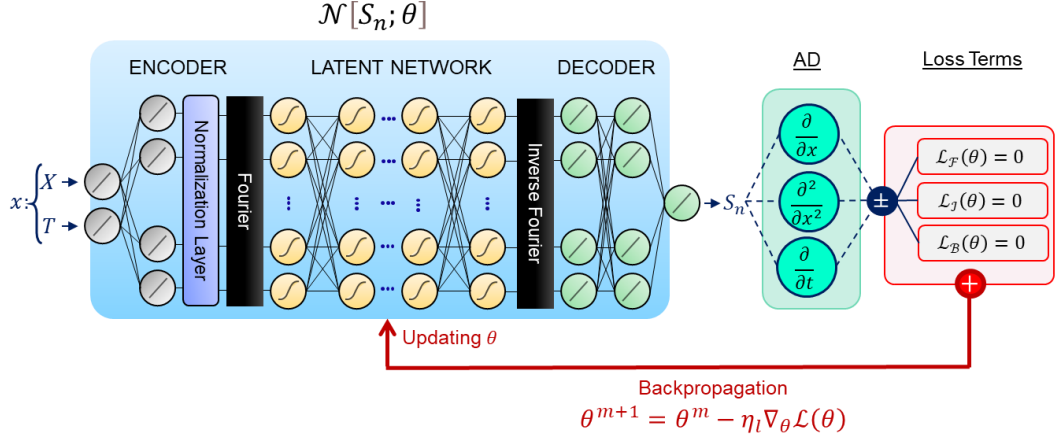

*Figure 3: The architecture of the PINN based on the $S_n(X, T)$ formulation. Based on the applied PDE form, the input can also be X and Y for the XY PDE, and only Z for the Z PDE. The neurons in the latent network were activated by the tanh activation function. Through the utilization of the back-propagation approach, the network's trainable parameters (θ) were dynamically updated throughout the training process. The primary objective was to effectively minimize the total loss term at the chosen collocation points (x).*

## 4- Results
### 4-1- PINNs solutions

Results of the trained PINN solutions are presented in this section for cases A and B, using the three formulations introduced previously. The reported results from PINN correspond to the training after 4000 epochs. Figure 4 compares the saturation and recovery profiles for XT and XY formulations at all times. Both formulations could capture the recovery profile acceptably. However, the overall error in the XY formulation was lower than the XT formulation in all time periods.

Comparing the ET saturation profiles of different formulations (XT and XY in Figure 4, and Z formulation in Figure 5) shows that all PINN with PDE formulations capture the overall trend in the ET period. The most challenging aspect seems to be capturing the boundary conditions where it is seen that the saturations at $X = 0$ approach values less than one, and the saturations at the imbibition front tend to overshoot the FD solution. This difficulty causes a greater MAE in saturations at ET compared to LT. The solutions overlap better at central saturations and for saturation profiles that have advanced further into the core. The mentioned challenges at the boundaries were handled better by the Z formulation than XT and XY formulations.

For each case and formulation, the FD and PINN saturation solutions, the absolute errors between them, and PDE residuals are illustrated in Figure 6 in 2D plots vs. spatial locations and time (Y). A comparison of the saturation profiles (Figure 6a and b), and their corresponding absolute errors (Figure 6c) reveals that the most errors accumulated in the ET period, and at regions close to the flowing boundary (X=0). A similar behavior is also depictable from Figure 6d for the residuals of the PINN loss term, where there were optimization difficulties at points close to X=0.

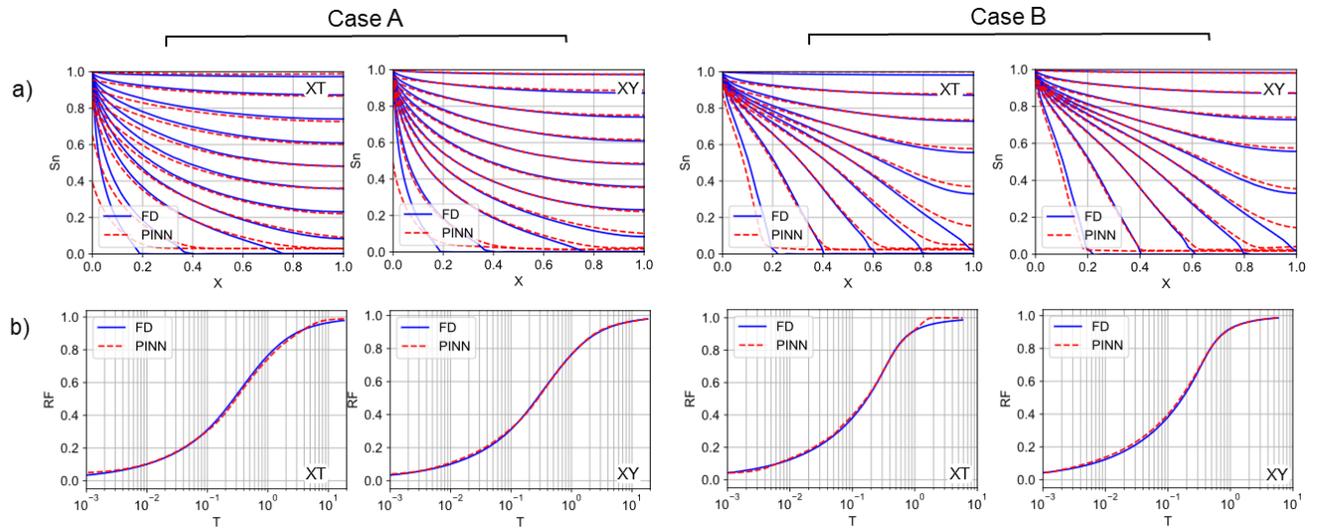

*Figure 4: The production profile for the XT and XY formulations in the entire COUCSI period. a) Saturation Profile, b) Recovery profile.*

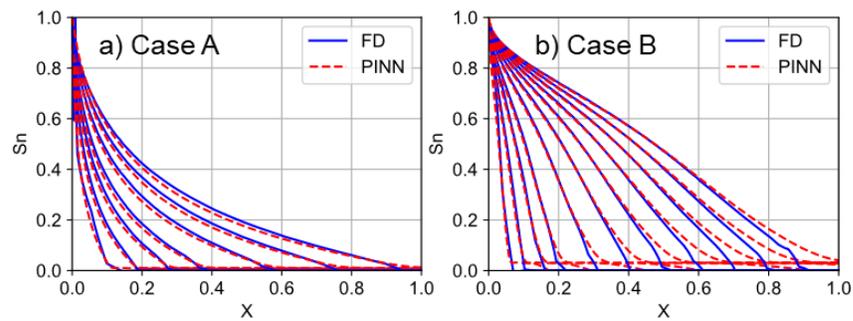

*Figure 5: Saturation profiles $S_n(X)$ obtained from Z formulation for the ET period of COUCSI. a) Case A, b) Case B.*

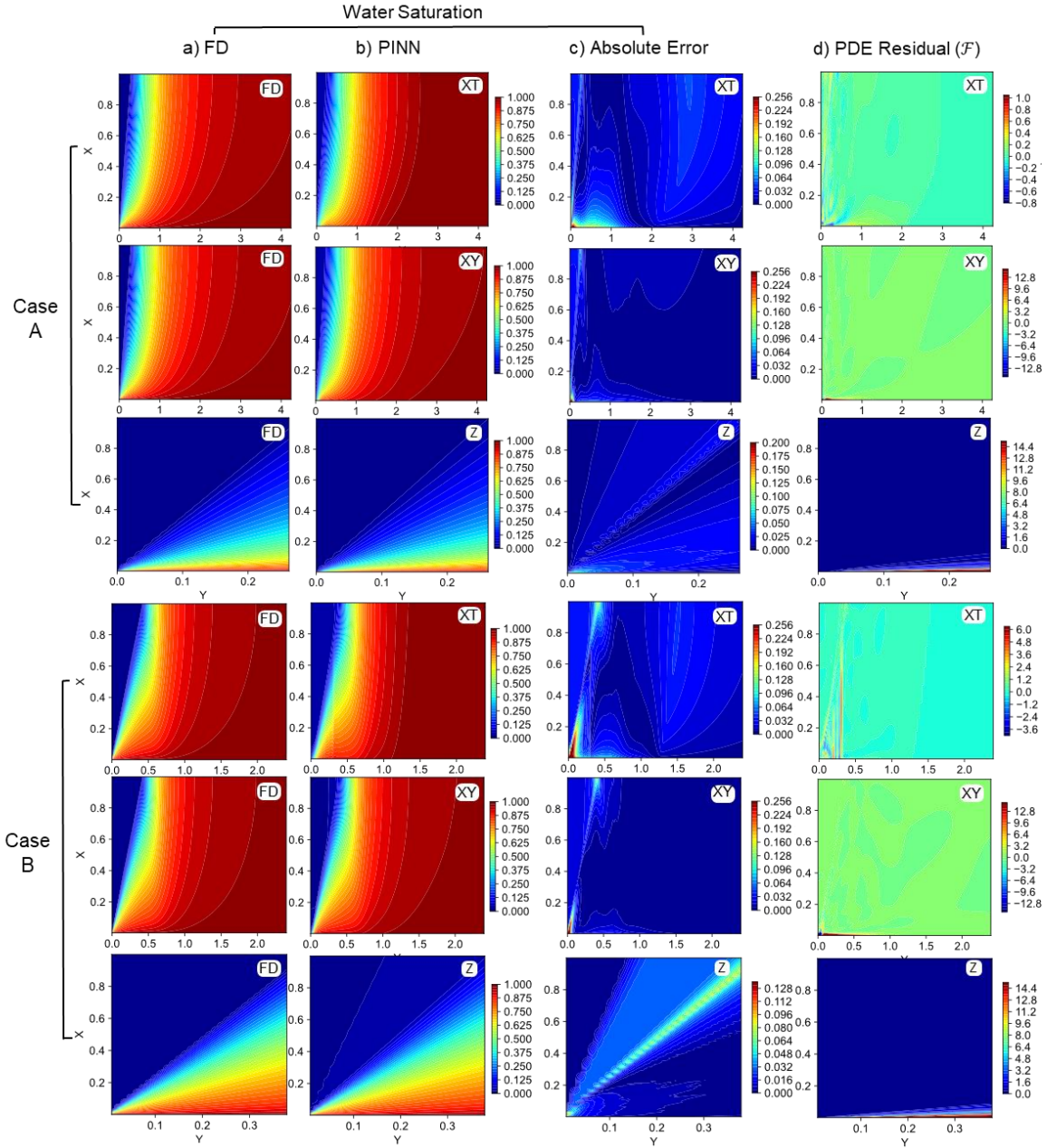

*Figure 6: The contour maps of FD versus PINNs solutions ($S_n$) for different formulations. The solutions for Z-formulation only show the ET period; a) baseline FD, b) PINNs, c) absolute error, d) PDE residuals (in each formulation, the scales are different, so the numbers are not directly comparable).*

A summary of the mean absolute error (MAE) values for PINN solutions (Saturation and Recovery factor) are shown in Figure 7. The values are separated based on ET, LT, or FT periods. The complete MAE values for the loss terms and solution outputs (saturation, capillary pressure, and recovery factor) are available in supplementary materials, Table S1. The MAEs are calculated by evaluating and comparing the PINN and FD solutions at the same collocation points. Recovery factor (RF) is defined as the average water saturation in the core at each time. All three formulations could capture the solutions of both COUCSI cases with reasonable accuracy. Figure 7 shows that the XY formulation performed superior in the FT period with the saturation MAE equal to 0.009 for both cases A and B, with roughly 50% less error compared to the XT formulation with MAE equal to 0.019 (case A) and 0.021 (case B). During the ET period, the Z formulation, with the $S_n$, the mean MAE of ~ 0.012 was superior to both other formulations (mean MAE ~0.030 for XT and ~0.014 for XY) in both cases. The same trends in performance were seen in Table S1 regarding the MAE of capillary pressure and recovery factor. Overall, there was a more significant error in ET as compared to LT, which can be attributed to PINN's inability to effectively minimize the loss term in points located near the flowing boundary, as it was discussed previously.

One crucial factor that influenced the efficacy of PINN was the approach used for sampling collocation points. The comprehensive analysis is presented in the supplementary materials, A4. Our findings indicate that selecting points based on constant DY intervals led to a notable decrease in the required data points. Furthermore, when the same number of data points were used, the model based on constant DY intervals exhibited superior performance. It is worth noting that the COUCSI problem exhibits a more uniform flow rate on the Y coordinate, than the T coordinate. Another crucial point to consider is that the XY and Z formulations performed better with lower numbers of collocation points (See Figure S2), which is highly beneficial for the rapid solution of the COUCSI problem in practical scenarios.

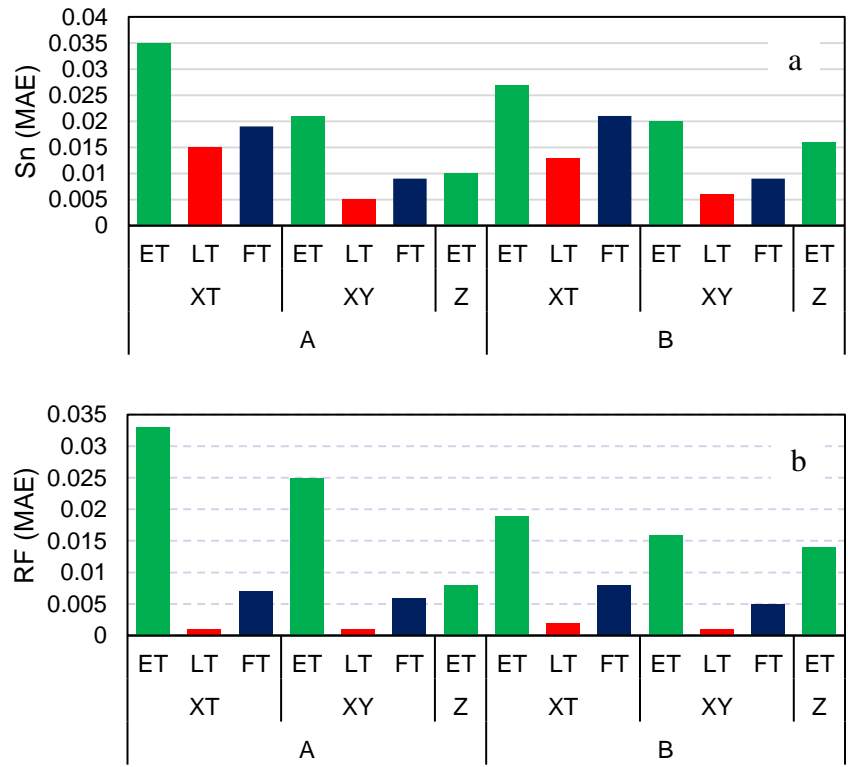

*Figure 7: The final error (MAE) of PINN predictions for all three formulations, separated by early-time (ET), late-time (LT), and full-time (FT) periods. a) Sn, b) RF.*

Figure 8 shows how the total loss term ($\mathcal{L}_t$) and saturation MAE of the networks evolve with training. Note that the predicted saturations are obtained from training the networks to satisfy the loss terms of the PDEs, initial conditions, and boundary conditions, but not data directly. The error is based on comparing predicted saturations with the FD saturation values. After 500-1000 epochs, the PINN solution errors stabilized very close to the FD values, even with low learning rates; after that, we continued the training for fine-tuning the solutions. The solutions could have been achieved faster by choosing a higher starting learning rate.

Nonetheless, when evaluating the effectiveness of PINN across various formulations, it was found that the XY and Z formulations achieved convergence towards the correct solutions in fewer epochs than the XT formulation. The comparison of run-time for PINN in supplementary materials, A6, also indicates that the XY and Z formulations provided quicker solutions as compared to the XT formulation (See Figure S5). Moreover, all the models demonstrated a satisfactory correlation between the total loss term value and the saturation profile's MAE (as depicted in Figure S1), signifying the dependability of the PINN solution in the absence of any knowledge pertaining to the true solutions.

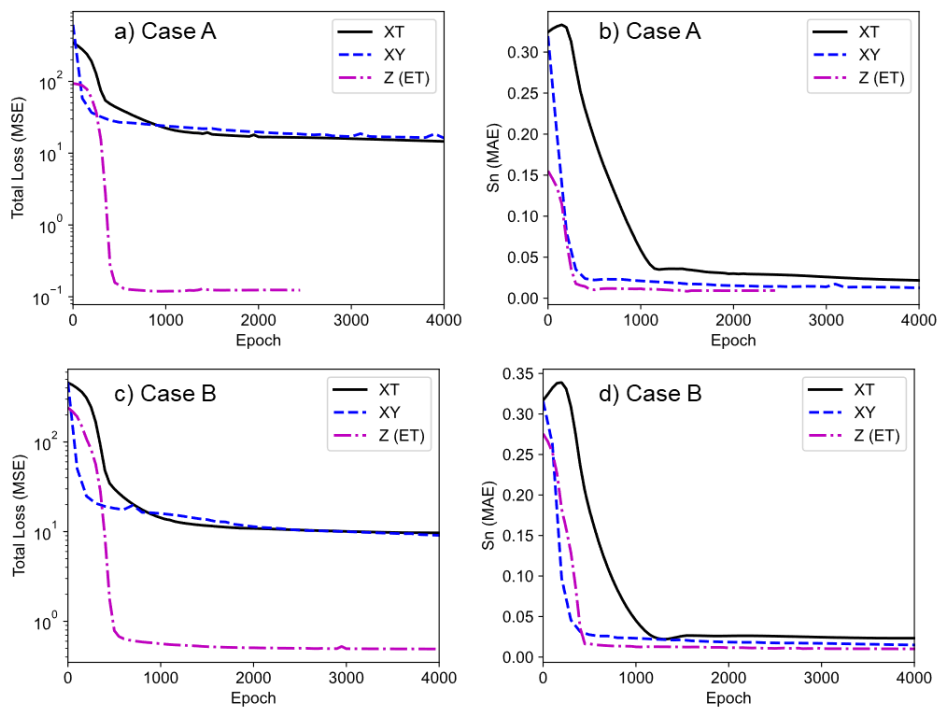

*Figure 8: The trend of the total loss term and Sn MAE versus epochs during the training for different PDE formulations. a-b) Case A, c-d) Case B.*

### 4-2- Similarity properties of COUCSI

In Figure 9, spatial saturation profiles ($S_n$ vs. $Z$) and recovery profiles ($RF$ vs. $Z$) are shown in the ET period. Figure 9a shows the saturation profiles plotted versus the similarity variable Z for the different timesteps. Ideally, we expect all the profiles to overlap because of the self-similarity of COUCSI at ET. In the Z formulation, all the profiles are matched. By design, it preserves self-similarity as the PINN solution is only a function of $Z$ (different ranges of $Z$ follow from considering different times). The XT and XY formulations failed in giving proper self-similar solutions at the beginning times of flow: the saturations of a given $Z$-value on some curves was lower than the correct values.

Figure 9b checks if recovery predicted from the PINN solutions followed proportionality with the square root of time. Semi-analytical solutions for 1D COUCSI by McWhorter & Sunada [15] have proven that recovery at ET follows a linear trend versus the square root of time (linearity with $Y$). The Z formulation produced a straight line for recovery against *Y* during ET, where the slope matched well, but not precisely, with the FD solution. The XY formulation followed the FD line well but was not a perfectly straight line; the slope rather slightly increased with Y. This was the same for the XT formulation, where the PINN solution more notably failed to give zero recovery at zero time but corresponded well with the FD solution at higher values of Y.

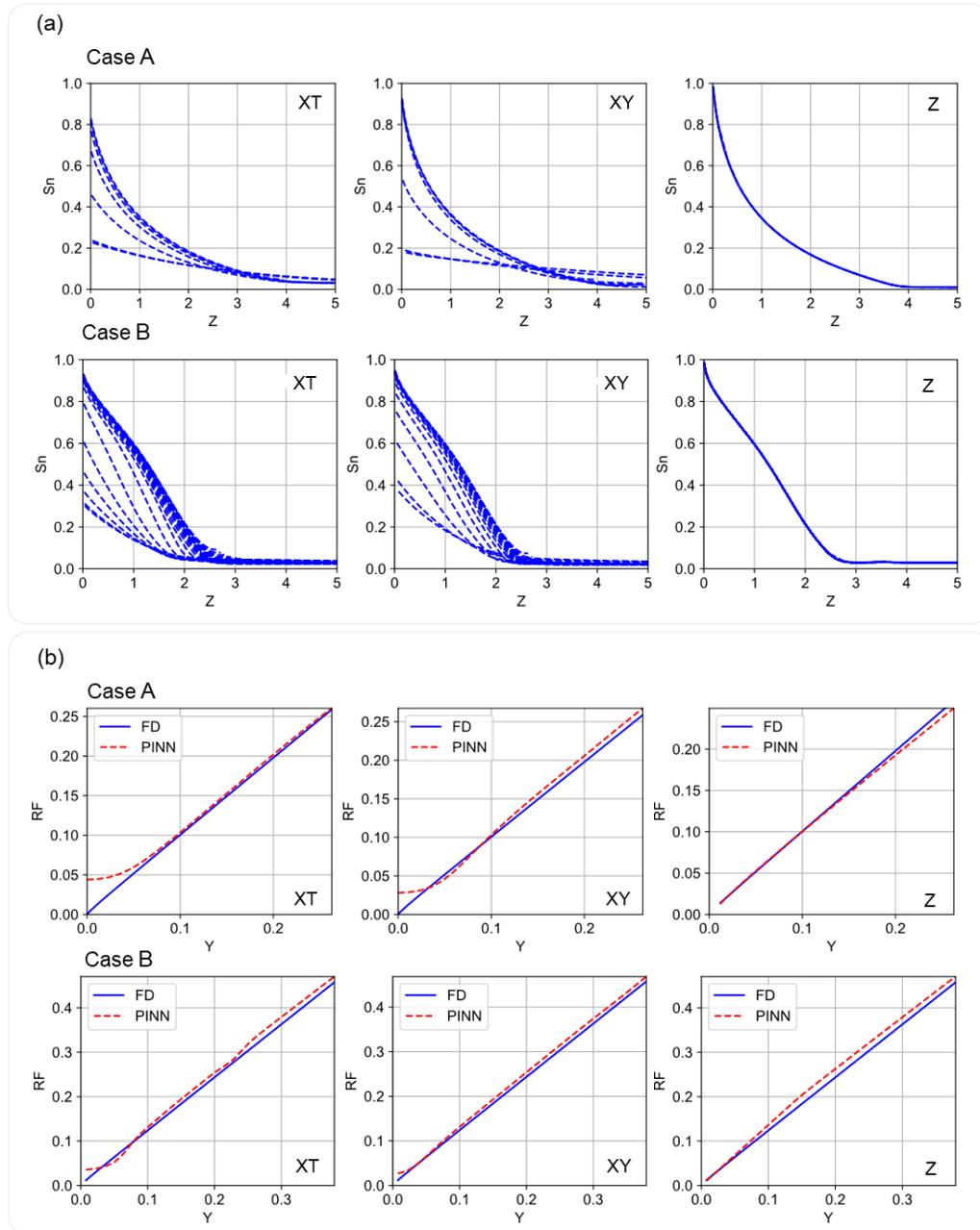

*Figure 9: The performance of PINNs with different PDE formulations (XT, XY, Z) for the ET period of COUCSI. a) Saturation profiles vs. similarity variable (Z), b) Recovery factor versus square root of time (Y) for FD and PINN solutions.*

## 4-3- Comparison of Total Variations (TV)

The TV behavior of different solutions in the ET period are compared in Figure 10 as a function of scaled time T. The curves show that the *XT* formulation performed weakly in preserving the TV property of the model through time and that the value stayed far below 1. XY formulation could preserve the TV at a higher value, while the Z formulation managed to stay fairly constant close to one. In the XT solution, TV increased with T, indicating the inlet saturation increased with time. As this saturation was more fixed in the other formulations, TV did not change with time in the beginning for those cases. The TV decreased towards the end in all formulations because saturations were predicted to reach the closed boundary too early. This was a bigger problem in Case A than B and more severe for the XT and XY formulations than the Z formulation. A majority of the deviations of TV from 1 originate from the weaknesses of the PINN in sufficient handling of the $\mathcal{I}$ and $\mathcal{B}_1$ terms (see Table S1).

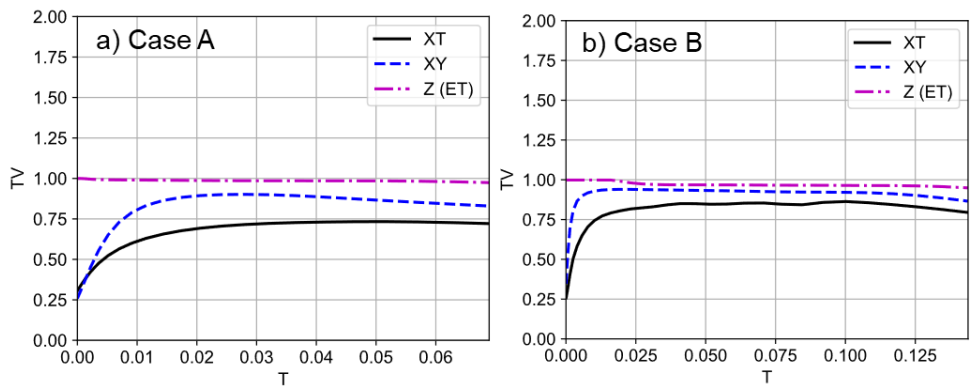

*Figure 10: The calculated total variation (TV) from the PINN solutions with different PDE formulations (XT, XY, Z) for the ET period of COUCSI. a) Case A, b) Case B.*

## 4-4- Comparison of loss landscapes

To see the impact of changing formulations in PINNs, we compared the loss landscape for all three formulations in Figure 11. The loss landscape visualizes the optimization path of neural networks by employing dimensionality reduction techniques, allowing for a more intuitive and efficient analysis of the network's parameter adjustments during the optimization process [43]. This insight can aid in understanding the network's behavior, PINN's overall model performance (the mathematical details are provided in section A3 of the supplementary materials). It should be noted that the central point of each landscape is related to the trained PINN model. Figure 11a shows a chaotic, non-convex landscape with an approximately ambiguous global minima, while XY formulation clearly shows a smooth surface with a trivial global minima. Z formulation also showed a smooth loss landscape with a clear global minima. The comparison of the loss landscapes reveals that changing the variables of the PDE from XT to XY and Z formulations could significantly enhance the convexity and smoothness of the loss landscape and provide training trajectories that are more explicit and dependable. This enhancement results in a reduction in the number of required collocation points (refer to section A4 of supplementary materials), the necessary network size (refer to section A5 of the supplementary materials), as well as the run-time of the model (refer to section A6 of supplementary materials).

16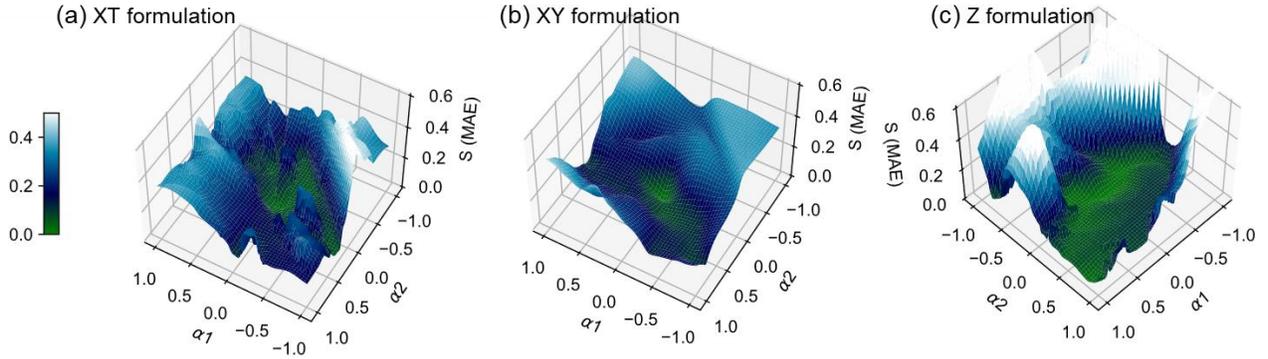

*Figure 11: 3D loss landscape generated by linear mixing of two different initial directions with the trained model parameters. Here, $\alpha_1$ and $\alpha_2$ are direction vectors that is varied in the range of 0 to 1 (further details in the section A3 of the supplementary materials). a) XT, b) XY, c) Z.*

### 4-5- Validation with experimental tests

Finally, we tested PINNs (XY formulation) to regenerate three experimental cases and compared the PINN solutions to the FD-based results. The experimental results were reported by Fischer et al. [44] for different 1D COUCSI cases with different viscosity ratios. The interpreted saturation functions were provided by Andersen [8]. The $\Lambda_n$ curves are visualized in Figure 12a and the results of the forward simulation in Figure 12b. In all three cases, PINNs successfully regenerated the recovery profile of the imbibition process, closely matching the results obtained using FD and the experimental data.

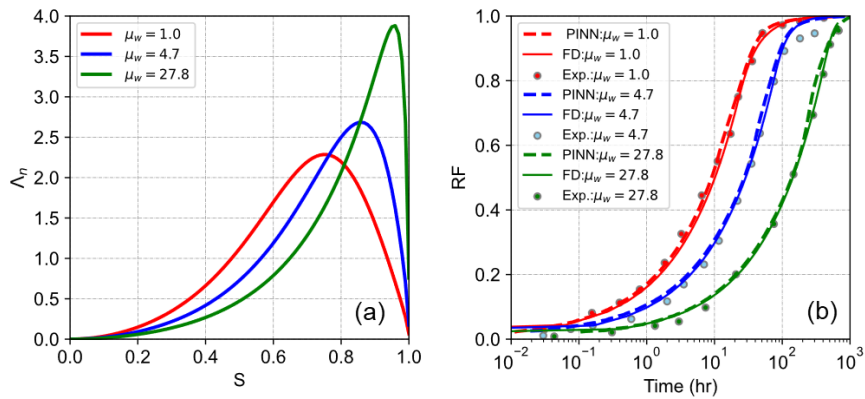

*Figure 12: A comparison of PINN and FD (IORCoreSim) simulation of the experimental 1D COUCSI tests reported by Fischer et al. [44]. a) the applied CDC curves, b) the semi-log recovery plot.*

# 5- Discussion

This study focused on PINN-based solving of the two-phase 1D COUCSI phenomena with different equivalent PDE formulations obtained from the Change-of-Variables technique. PINN could successfully solve all the PDE formulations with acceptable errors. Changing the PDE variables in accordance with the physical properties of the system, such as the square root of time flow, showed to have the potential to enhance the performance of PINN in various ways, including accuracy of solutions and computational efficiency of the models.

A severe problem in many of the PINN models during the training was that the optimizer could quickly lose track of the actual solution and converge to the trivial solution with $S_n(x,t) = 1$ in all spatial and temporal values. We found that the problem can be tackled by choosing approximately equal weight factors for different terms in the loss function and choosing enough resolution of the collocation points, especially in points close to the initial/boundary regions. A practical problem with the $Z$ formulation was that at $T = 0$, $Z_{max} = \frac{1}{Y} \to \infty$ and the interval with non-trivial solutions becomes very small compared to the full interval. We tackled this problem by adding a small constant to the denominator when selecting collocation points.

The total variation TV is useful for evaluating mathematical systems and their numerical schemes. In this work, we utilized this concept to track the PINN results' validity for the different cases. In the problems studied in this work, the TV should stay constant, equal to one, at ET and then decline towards zero. The preservation was satisfied to a reasonable level in $Z$ formulation. For the XT (and, to a less extent, XY) formulation, TV varied more dynamically, starting and ending at lower values than the others, while for all the formulations, TV declined towards the end. The difference from one was related to lower saturation at the inlet than the mathematical boundary condition. It stayed constant for Z formulation, while for XT (and XY), it changed dynamically. The late decline was related to PINN saturations reaching the no-flow boundary too soon. We believe the TV property can be better maintained by inclusion as an additional loss term in the training process or by better capturing the boundary condition loss terms. There are generally two different approaches for defining this term: 1) the *TV* may be calculated analytically using automatic differentiation; this approach should be more accurate; however, it may add significant computational load to the optimizer. 2) the more efficient approach is the definition of the problem in a discretized approach where it still has acceptable accuracy but takes much lower computational costs.

In addition to capturing the PDE, IC, and BC terms, we note that the PINN should capture two important transitions: the finite saturation gradient at the front jumps to zero ahead of the front. As seen in Figures 5 to 7, the PINN's main challenge was capturing the abrupt change in gradient at the front and the self-similarity of the ET solution when two variables were involved in the system formulation. In case B, the gradient was steeper at the front, and hence the change to zero gradients ahead of the front was more abrupt, which is believed to explain the incredible difficulty of training the solutions to this case.

We could show the feasibility of applying a pure PINN-based approach for solving benchmark core-scale phenomena with different formulations. The merit of using different formulations was that we could show the potentials of using the Change-of-Variables technique for improving the performance of PINNs in terms of collocation point selection, network complexity and by reducing the complexity of PDEs and simplify complicated differentiations. We believe the approach can be extended to other engineering processes. However, we see a potential to achieve lower error values and reduce the computational load during training. To facilitate the comparisons, we employed a basic version of PINNs, essentially vanilla PINNs. Nevertheless, for future studies, more recent advancements like adaptive activation functions and self-adaptive loss weights could be implemented to enhance the accuracy of the solutions. Given its relatively short run-time compared to numerical simulations, PINN-based solutions are a promising approach for obtaining quick approximations of forward problems. However, PINNs should be clearly advantageous for inverse problems, where the iterative nature of numerical simulation methods significantly increases computational time. Our future work will explore the use of PINNs for inverse solutions in porous media flow processes.

## 6- Conclusions

In this work, we studied the performance of PINNs in solving the two-phase 1D COUCSI problem. Using Change-of-Variables, the system was expressed in three equivalent PDE formulations (XT, XY, and Z), and we compared the resulting performance of the PINNs during early- or late-time for two simulation cases A and B. The following conclusions were obtained:

- The PINN solver could acceptably solve the 1D COUCSI problem, as shown for two different cases in early and late time flow regimes, with the MAE of saturations around 0.01 for the ET and 0.03 for the full imbibition time. The error was typically least for Z formulation (only ET), followed by XY, then XT. The errors were typically located at the inlet, at the imbibition front and at the closed boundary when the front arrived there.

- Total Variation of saturation was used to evaluate the saturation profile solutions with time. A difference from 1 at ET indicates that the solution does not meet the boundary and initial conditions properly (or has oscillations, which was not encountered).

- The Change-of-Variables technique helped in transforming the coordinates of the imbibition problem into a neater space which had lower prediction errors and was closer to the physical properties of the COUCSI problem, such as self-similar saturation profiles, TV preservation and the square root of time recovery behavior in the ET region. Especially, the Z formulation, by definition, obtained self-similar profiles, square root of time recovery, and constant TV. In contrast, XT and XY formulations obtained profiles that did not fully overlap at ET, recoveries that only approximately followed the square root of time and were not identical to zero at time zero, and TVs that varied with time. XY formulation, however, performed much better than XT from these perspectives. The TV values at ET were closer to 1 for Z, then XY, then XT formulations indicating a difference in how well the initial and boundary conditions were captured. The comparison of loss landscapes also revealed that the XY and Z formulations had a higher degree of convexity as compared to the XT formulation.

- By Change-of-Variables, fewer collocation points were required depending on which formulation and variable sampling procedure was used. Both for XT and XY systems, fewer temporal points are needed if they are selected based on constant DY (fixed square root of time difference) compared to constant DT (fixed time difference) and lower errors are obtained. The constant DY sampling focuses on both the early and late times. At a given number of temporal points, the XY formulation had less error than the XT formulation. Also, The Change-of-Variable technique could help reduce the network size.

- The run-time of PINN to achieve a saturation MAE of 0.03 was shorter than that of MRST, while it was longer if one aimed to attain lower MAE values. Overall, it can be inferred that the solutions provided by PINN are computationally efficient if one requires a fast estimation of the accurate solutions.

For future studies, PINNs-based calculations could prove highly advantageous for inverse problems in porous media, especially compared to traditional numerical methods that involve time-consuming iterative calculations.

# Supporting Information

The Supporting Information is available at [link].

**A1:** Mathematical derivation of the scaled representation of the PDE corresponding to the COUCSI problem; **A2:** Complementary details regarding the results of PINN Training, including the values of different loss terms for different scenarios of the study; **A3:** Complementary (mathematical) details regarding the visualization of loss landscapes; **A4:** Evaluation of the different collocation points sampling schemes, including the temporal distribution and the number of the points; **A5:** Sensitivity analysis on the impact of network size on the accuracy of the solutions; **A6:** Comparison of the computational costs of PINNs and finite-difference based numerical simulations.

# Acknowledgments

The authors acknowledge the Research Council of Norway and the industry partners of NCS2030 – RCN project number 331644 – for their support. Also, we appreciate Validé AS for partially granting this project under the Plogen program.

# Nomenclature

### Symbols and Abbreviations

ET  Early-time,

FT  Full-time,

$f_w$  Fractional water flow function,

$J$  Scaled capillary pressure,

$K$  Rock permeability, m$^2$

$L$  Core length, m

LT  Late-time,

$n_i$  Corey exponent,

$p_i$  Phase pressure, Pa

$RF$  Recovery factor,

$s_w$  Water saturation,

$S$  Water saturation normalized over mobile saturation range,

$t$  Time, s

$T$  Normalized time,

TV Total variation,

$u$  Darcy velocity, m/s

$X$  Scaled length,

$Y$  Square root of the normalized time,

$Z$  X/Y,

**Greek Letters**

$\Lambda$  Dimensionless capillary diffusion coefficient,

$\Lambda_n$ Normalized capillary diffusion coefficient with mean 1,

$\overline{\Lambda}$  Mean of dimensionless capillary diffusion coefficient,

$\tau$  Recovery time scale, s

$\mu_i$  Phase viscosity, Pa s

$\phi$  Rock porosity,

$\sigma$  Interfacial tension, N/m